\documentclass{optica-article}

\journal{opticajournal} 

\articletype{Research Article}

\usepackage{breqn}
\begin{document}

\title{High Forward Thrust Metasurface Beam-Riding Sail}

\author{Prateek R. Srivastava,\authormark{1} Apratim Majumdar,\authormark{2} Rajesh Menon,\authormark{2} and Grover A. Swartzlander, Jr.\authormark{1,*}}

\address{\authormark{1}Chester F. Carlson Center for Imaging Science, Rochester Institute of Technology, Rochester, NY 14623, USA\\
\authormark{2}Department of Electrical and Computer Engineering, The University of Utah, Salt Lake City, UT 84112, USA}

\email{\authormark{*}grover.swartzlander@gmail.com} 


\begin{abstract} 
The radiation pressure force and torque on a one-dimensional bi-grating composed of a 
$\mathrm{Si-SiO_2}$ high contrast binary metagrating is analyzed for the purpose of stable beam riding whereupon a high power laser having an expanding Gaussian irradiance distribution propels the grating in outer space, free from gravitational forces.  The binary metagrating structure has been simultaneously optimized to afford high forward thrust, and corrective restoring forces and torques 
in the event of small linear and angular disturbances. We demonstrate that stability may be enhanced at the expense of forward thrust.
The validity of our metamaterial findings is reinforced owing to good agreements between finite-difference time-domain and finite element numerical methods.  To reduce mass and enhance forward acceleration this laser-driven sail was designed to be free of a stabilizing boom.
\end{abstract}

\section{Introduction}

Optical tweezers or single-beam optical traps are one of the most powerful means of contactless manipulation of nanoscopic/microscopic objects with a very diverse set of applications ranging from biology\cite{Dholakia2011} to quantum optomechanics\cite{Bhattacharya2017}. But the promise of optical tweezers to manipulate macroscopic objects of meter-scale has been held in check by the need for a tightly-focused beam that creates a gradient force in a very limited volume and effective distance. By engineering the scattering properties of the objects and the shape of the beam, the dynamics of macroscopic objects may be controlled and used for successful trapping, manipulation, levitation, and even propulsion without the need for a tightly focused beam. Most recently, NASA and Breakthrough Starshot Initiative aim to leverage these radiation pressure forces to propel a meter-scale light sail to the nearest stars with relativistic speeds. Nearly all the stable geometric designs proposed so far in the literature are subprime in three senses: (a) the ability of a conical/spherical/concave/convex sail to maintain its shape is questionable (b) a flat diffractive/nanophotonic design demands an undesirable mast to offset the center of mass away from the sail to achieve stability \cite{Srivastava2019} and (c) all the designs sacrifice on thrust force for levitation/propulsion to enable a restoring force for stability.

In this work, we demonstrate the stability of a bi-grating lightsail comprised of subwavelength unit-cells of Si/$\mathrm{SiO_2}$. The unit-cell geometry is optimized to engineer diffraction efficiencies such that the sail is stable against linear and rotational perturbation without any offset between the center of mass and the sail while simultaneously achieving maximum forward thrust. This work is different from \cite{Atwater2018} in the sense: (1) forward thrust is also an objective of optimization (2) a stability basin is compared between two different forward thrusts that quantify the trade-off between forward thrust and stable initial conditions. What's more, the forces and torque are validated using FEM (MEEP) and FDTD (COMSOL) methods. The two-dimensional dynamical analysis and electromagnetic response presented in this work may be extended into three dimensions using the scheme described in \cite{Srivastava2020,Kumar2021}.

\section{Theory}
\begin{figure} 
\centering\includegraphics[width=0.8\linewidth]{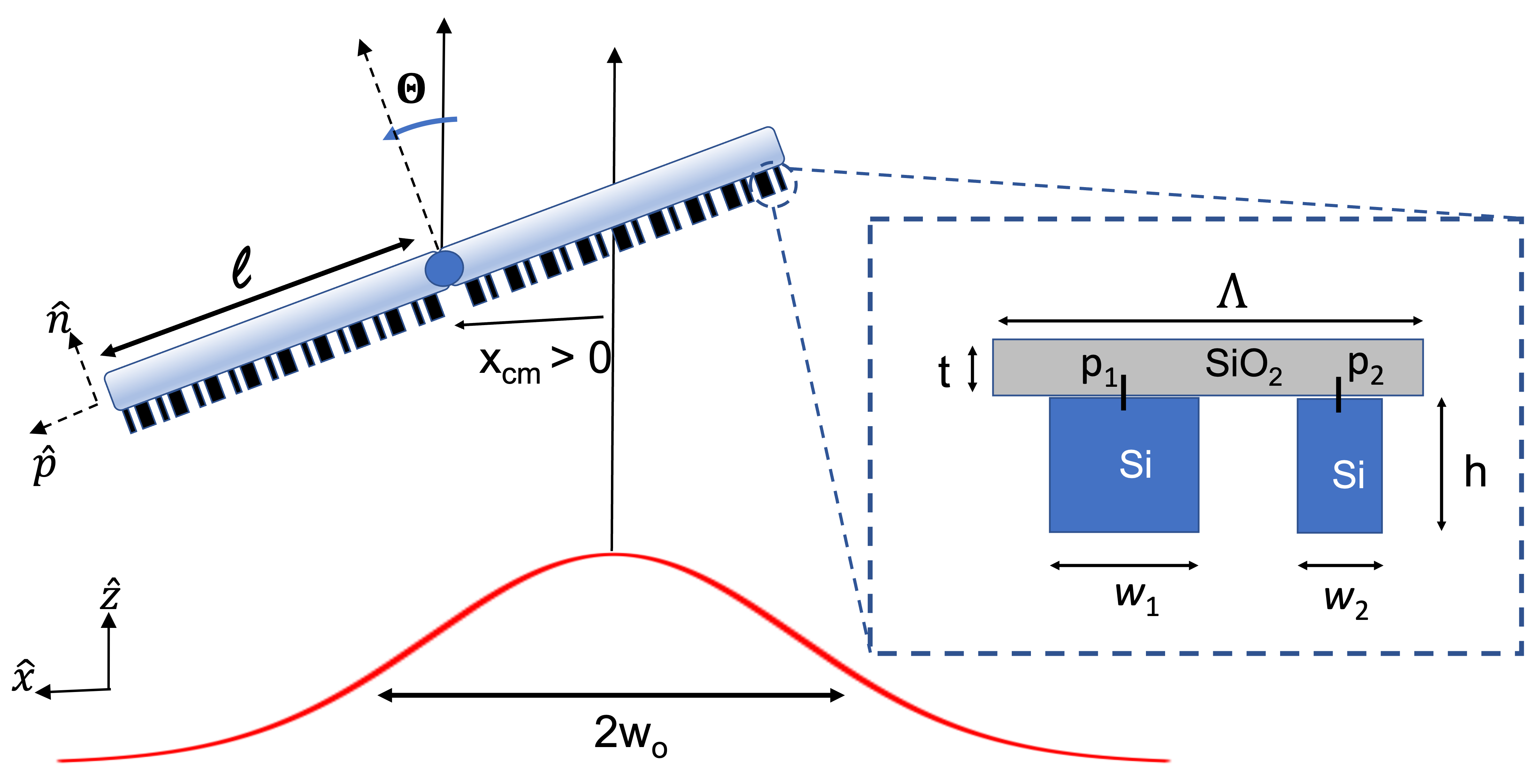}
\caption[]{High contrast metasurface bi-grating configuration propelled by a Gaussian beam. The bi-grating is comprised of unit-cells with geometric parameters shown in the inset. $p_{1,2}$ is measured from the center of the unit-cell.}
\label{fig:beamridingbi-grating}
\end{figure}
Consider the structure shown in Fig.\ref{fig:beamridingbi-grating} composed of two different panels $L$ (left) and $R$ (right) of length $\ell$. Each panel is comprised of an artificially designed sub-wavelength periodic lattice whose properties primarily arise from the design and distribution of meta-atoms or unit cells. Inspired by the principles of form birefringence and effective medium theory, the unit-cell of choice in this work is a ridge-width-modulated high contrast grating with $\mathrm{Si}$ nano-pillars on a low index $\mathrm{SiO_2}$ substrate. As shown in Fig.\ref{fig:beamridingbi-grating} (inset) is a multi-layer subwavlength binary unit-cell of period $\Lambda$. High-index Si nanopillars of height $h_1=h_2$ are deposited on low-index $\mathrm{SiO_2}$ substrate of thickness $t$. The width and position of the nanopillar is $w_{1,2}$ and $p_{1,2}$ respectively. There exist multiple benefits to using this subwavelength binary design: (a) this design has been shown to offer stable levitation \cite{Ilic2019}, (b) optimal thermal management via radiative cooling \cite{Salary2020}, and (c) the feature sizes may be easily realized with existing e-beam lithography technology

The bi-grating may levitate or propel when illuminated owing to radiation pressure forces. To model these forces, it
is convenient to consider two different frames of
reference: $(\hat{x}, \hat{z})$ and $(\hat{p}, \hat{n})$
attached to the laser and bi-grating respectively. The  bi-grating has a total mass $M$ (including a payload) 
and the center of the bi-grating coincides with 
 the center of mass of the system at $x_{cm}$ such that there is no offset between the two. The two coordinate spaces are related to each other as $p = (x - x_{cm})/\cos\Theta$ where $\Theta 
= -\theta_i$ is the attitude of the sail rotated about 
$\hat{y}$ and $\theta_{i}$ is the angle of the beam 
incident on the bi-grating.

The meta-atoms offer exceptional control of wavefront and scattering response near their resonance, which may be engineered for desired force and torque on the bi-grating structure. Assuming a non-relativistic non-spinning bi-grating, the incident and scattered wavelengths are equal in the reference frame of the structure, and thus, the respective wave vectors may be expressed 
$\vec{k}_{i}= k\hat{n}$ and 
$\vec{k}_{m}^{L,R}= k (\cos \theta_m^{L,R} \; \hat{n} + \sin\theta_m^{L,R} \; \hat{p})$, where $k = 2\pi/\lambda$.  The diffraction angles are governed by the grating equation,
\begin{equation}
 \sin\theta^{L,R}_{t_{m}}= -\sin\theta^{L,R}_{r_{m}} = m\mathbb{K}^{L,R} - \sin\Theta
\label{eq:diffractionEqnfor2panels}
\end{equation}
\noindent where and $\mathbb{K}^{L,R}$ is the grating momentum such that $\mathbb{K}^{L}=-\mathbb{K}^R$ and $(\theta^{L,R}_{t_{m}}$ , $\theta^{L,R}_{r_{m}})$ are the diffraction angles for the $m^{\mathrm{th}}$ (transmitted, reflected) orders. The change in photon momentum $\Delta\Vec{k}^{L,R}=(\Vec{k}_i-\Vec{k}_m^{L,R})/k_i$ in the reference frame of unit-cell may now be expressed 
\begin{equation}
    \begin{split}
        \Delta\Vec{k}^{L,R}(\Theta) = -\sin\Theta \ \hat{p} + \cos\Theta \ \hat{n}  -  \sum_{m=-1}^{m=1} &(\eta^{L,R}_{r_{m}}\sin\theta^{L,R}_{r_{m}} +  \eta^{L,R}_{{t_{m}}} \sin\theta^{L,R}_{t_{m}} )\hat{p} \\
        &- (\eta^{L,R}_{r_{m}} \cos\theta^{L,R}_{r_{m}} + \eta^{L,R}_{{t_{m}}} \cos\theta^{L,R}_{t_{m}}) \hat{n}
    \end{split}
    \label{eq:deltaKtheta}
\end{equation}
\noindent where $\eta^{L,R}_{r_m, t_m}$ is the diffraction efficiency of $m^{\mathrm{th}}$ reflected($r$) or transmitted($t$) modes from corresponding panels ($L,R$). 
Alternatively, the change in photon momentum may also be expressed 
\begin{equation}
    \Delta\Vec{k}^{L,R}(\Theta) = \frac{1}{P_0/c}\left( \int_{\partial S} T_{ij} \cdot \hat{n} dA + \int_{\partial S} T_{ij} \cdot \hat{p} dA \right)
\end{equation}
where $T_{ij}$ the Maxwell Stress Tensor expressed as.
\begin{equation}
 T_{ij} = \epsilon_0(E_iE_j - \frac{1}{2}|\mathbf{E}|^2 \delta_{ij}) + \frac{1}{\mu_0}(B_iB_j - \frac{1}{2}|\mathbf{B}|^2 \delta_{ij})
\end{equation}
where $E$ and $B$ are electric and magnetic field respectively, $mu_0$ and $epsilon_0$ are permeability and permittivity of free space and $\delta_{ij}$ is the kronecker delta function. 

The bi-grating may experience force and torque when illuminated by a laser beam of peak power $P_0$, wavelength $\lambda<<L$, characteristic beam-width $2w_0$, and a Gaussian irradiance distribution 
\begin{equation}
    I(p) = \frac{P_0}{2w_0^2\sqrt{\pi/2}}\cos\Theta\exp\left( -2\frac{\left( p\cos\Theta + x_{cm}\right)^2}{w_0^2}\right)
\end{equation}
and the force $\Vec{F}^{L,R}$ and torque $\vec{N}^{L,R}$ of each panel may now be expressed
\begin{subequations}
\begin{equation}
 \Vec{F}^{L}(p,\Theta) = \int_{-\ell}^0 \frac{I(p)}{c} \Delta\Vec{k}^{L} dp \quad \& \quad \Vec{F}^{R}(p,\Theta) = \int_{0}^{\ell} \frac{I(p)}{c} \Delta\Vec{k}^{L} dp
\end{equation}
\begin{equation}
\Vec{N}^L(p,\Theta) = \int_{-\ell}^0 p \hat{p} \times \frac{I(p)}{c} \Delta\Vec{k}^{L} dp \quad \& \quad \Vec{N}^R(p,\Theta) = \int_0^{\ell} p \hat{p} \times \frac{I(p)}{c} \Delta\Vec{k}^{R} dp
\end{equation}
\label{eq:ForceTorquenp}
\end{subequations}
\noindent such that the transverse ($F_x$) and longitudinal ($F_z$) forces in the reference frame of the beam may be expressed 
\begin{subequations}
\begin{equation}
 F_x = (F_p^{L} \cos\Theta + F_p^{R} \cos\Theta+ F_n^{L} \sin\Theta + F_n^{R} \sin\Theta)\hat{x}
\end{equation}
\begin{equation}
 F_z = (F_n^{L} \cos\Theta + F_n^{R}\cos\Theta - F_p^{L} \sin\Theta - F_p^{R} \sin\Theta )\hat{z}
\end{equation}
\end{subequations}
\noindent and the torque is the same in both the frames of reference.

The non-spinning two-dimensional system described above entails 3 degrees of freedom: translation along $\hat{x}$ and rotation about $\hat{y}$ while it is propelled along $\hat{z}$. We define a state vector $\mathbf{x} = [x, \theta]^T$ to analyze the stable transverse dynamics of the bi-grating system. In a close analogy of a oscillating spring system, the transverse dynamics of the bi-grating may be linearized and expressed as a set of ordinary differential equations (ODE) $\ddot{\mathbf{x}} = -K\mathbf{x}$, where \textit{K} is a Jacobian with \textit{stiffness} coefficients
\begin{equation}
 K = \begin{bmatrix}
k_1 & k_2 \\ 
k_3 & k_4 
\end{bmatrix} = \begin{bmatrix}
\frac{1}{M}\frac{\partial F_x}{\partial x} & \frac{1}{M}\frac{\partial F_x}{\partial \Theta}\\ 
\frac{1}{J_y}\frac{\partial N_y}{\partial x} & \frac{1}{J_y}\frac{\partial N_y}{\partial \Theta}
\end{bmatrix}_e
\label{eq:linearEqnjacobian}
\end{equation}
where the coefficients are evaluated in the close proximity of equilibrium point $\mathbf{x}_e = [0,0]^T$, implied by $e$. The metasurface is said to be marginally stable if $\mathrm{Im}(\mathbf{eig}(K)) = 0$ and $\mathrm{Re}(\mathbf{eig}(K))<0$ i.e., only real eigenvalues are allowed such that the frequencies of oscillation are real. In general, the system will exhibit two stable oscillation frequencies $\omega_{1,2} = \sqrt{-\mathrm{Re}(\mathbf{eig}(K))}/2\pi$ Hz. Note that, we add no restrictions on the nature of the stiffness coefficients, they can either be positive or negative.

We may conclude from Eq.\ref{eq:deltaKtheta} and Eq.\ref{eq:ForceTorquenp} that it is possible to engineer a unit-cell geometry that simultaneously high forward thrust and stability against small perturbation. However, the forces depends on $\eta$ and analytically characterizing the diffraction efficiencies is very difficult except for simpler geometries. For the aforementioned unit-cell, rigorous diffraction theory must be applied. We make use of an FDTD based package in Python called MEEP \cite{Oskooi2010meep} to solve for Maxwells Equation. The boundary conditions are assumed to be PML (Bloch periodic) along $\hat{n}$($\hat{p}$) . To account for fabrication constraints and numerical dispersion problem in FDTD, we limit our resolution to a step size of 20 nm or 50 pixels per micron. When an in-the plane polarizatied light of wavelength $\lambda$ is incident on the structure, the simulation is ran until the fields have decayed to $10^{-6}$ of their peak value. A similar FEA simulation is performed in COMSOL to validate the results from MEEP. The problem may now be formulated in terms of a multi-objective optimization problem i.e., for the set of variables $(\Lambda, h, p_{1,2}, w_{1,2})$ along with $\mathrm{Si (n=3.5220) /SiO_2 (n=1.4582)}$, we are seeking a sail design with with two figures of merit (FOM): (a) $\mathrm{FOM_1} = \mathrm{Im}(\mathbf{eig}(K))$ is minimized until it is 0 and (b) the forward thrust $\mathrm{F_z/F_0}$ is maximized, where $F_0 = 2P_0/c$.
The optimization is performed using a genetic algorithm called NSGA-II \cite{Deb2002} for the bi-grating-beam system for the following parameters: $L=1$, $w_0=0.5L$, $\lambda=1.2 \mu \mathrm{m}$, $M=1 \mathrm{gm}$, and $P_0=10$ kW.  

\begin{table}[t]
\centering
\caption{Optimized Geometries [in $\mu$m] (t=0.5 $\mu$m, $\lambda=1.2$ $\mu$m)}
\begin{tabular}{ccccccccccc}
\hline
  & $n_{\mathrm{Si}}$ & $n_{\mathrm{SiO_2}}$ &$\Lambda$ & h & $p_1$ & $w_1$ & $p_2$ & $w_2$ & $F_z/(P_0/c)$ \\
\hline
Unit-cell I & 3.5220 & 1.4582 & 2.24  & 0.78 &  -0.50& 0.80 & 0.46 & 0.56 & 120\%\\
Unit-cell II & 3.5220 & 1.4582 & 2.24  & 0.80 & -0.52 & 0.76 & 0.46 & 0.54 & 170\%\\
\hline
\end{tabular}
 \label{tab:geometricparameters}
\end{table}
  
\begin{figure}[b] 
\centering\includegraphics[width=\linewidth]{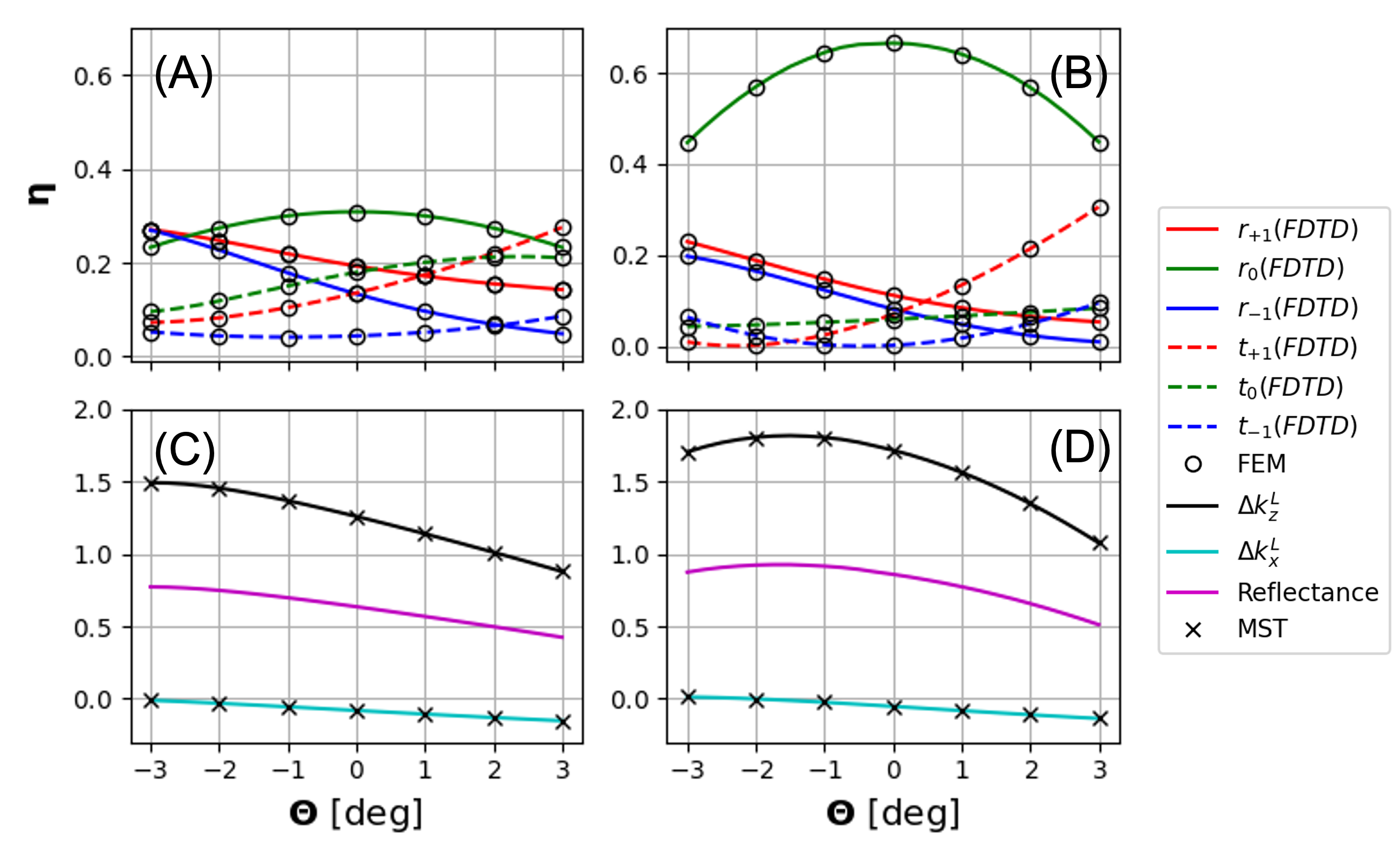}
\caption[]{The diffraction efficiencies of each orders for a sail tilted by $\Theta$. (A) and (B) correspond to the two structures with $F_z/(P_0/c)=1.26$ and $F_z/(P_0/c)=1.70$ respectively. Straight Line and Circles correspond to MEEP (FDTD) and COMSOL (FEM) results respectively.  }
\label{fig:etaVstheta}
\end{figure}

\section{Result \& Analysis}
Since, this is a multi-objective optimization problem, there is no single solution. Instead, there exist many solutions that are Pareto-optimal i.e., lie at the optimal trade-off between two competing objectives. From the multiple Pareto-optimal solutions that were achieved, selected some fabrication-friendly designs shown in Table \ref{tab:geometricparameters}. The diffraction efficiency of both the unit-cell geometries comprising the left panel is evaluated using MEEP(FDTD) and COMSOL(FEA) and is shown in Fig.\ref{fig:etaVstheta}(a) and (b) as the function of incident angle. An excellent agreement is evident between the two methods. The change in photon momentum $\delta \Vec{K}(\theta)$ is evaluated using both diffraction efficiency and Maxwell stress tensor and an excellent agreement between the two may be seen in Fig.\ref{fig:etaVstheta}(c) and (d). Clearly both the structure have very high reflectance that enables high forward thrust and hence $\Delta k^L_z $, whereas a non-zero  $\Delta k^L_x $ enables restoring stable force for non-equilibrium positions, as shown in Fig.\ref{fig:etaVstheta}(c) and (d). 

\begin{figure} 
\centering\includegraphics[width=0.9\linewidth]{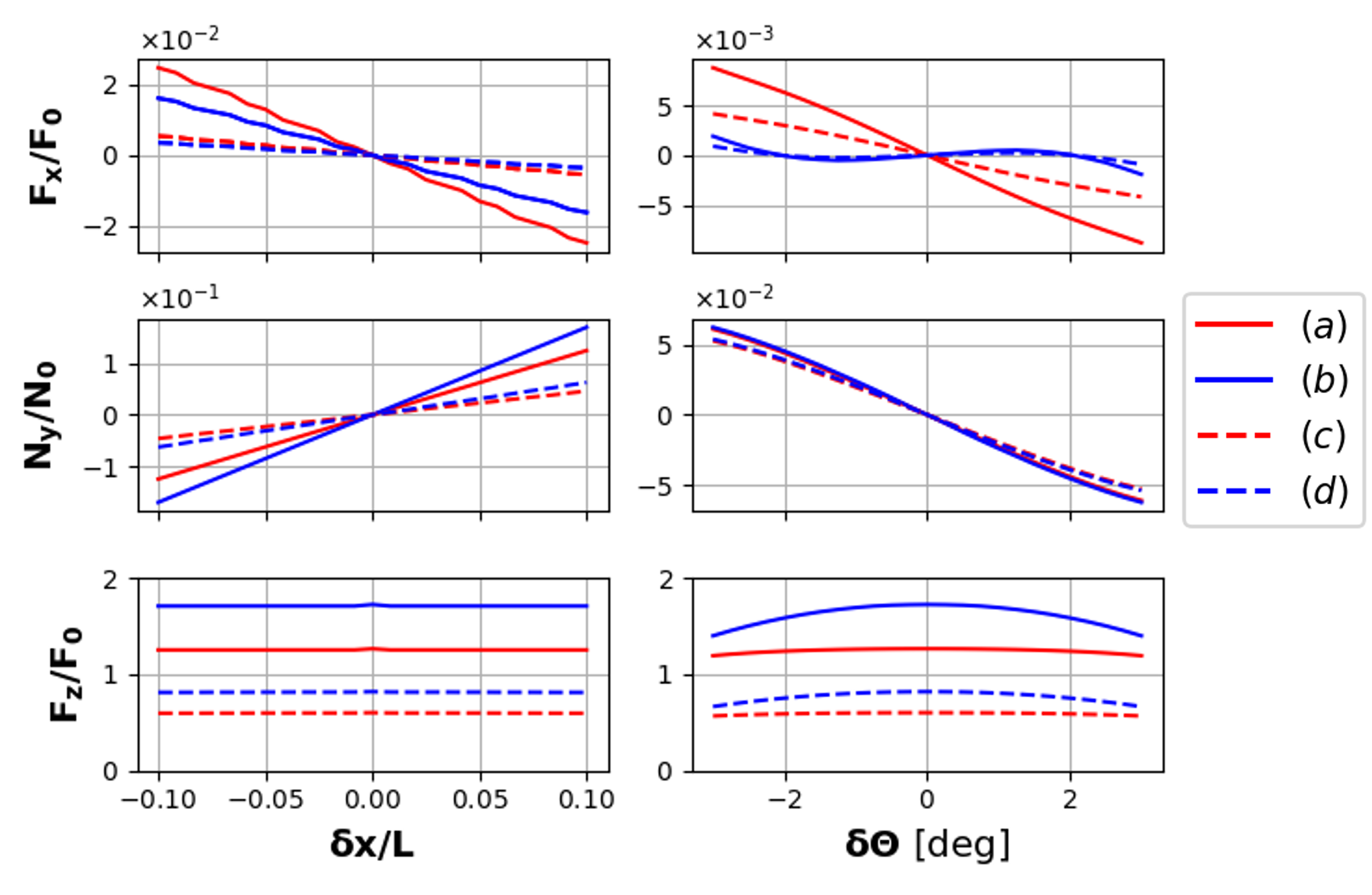}
\caption[]{Force and Torque as a function of displacement and rotational perturbation for the two sail unit-cell designs.}
\label{fig:Kmatrix}
\end{figure}

\begin{figure} 
\centering\includegraphics[width=\linewidth]{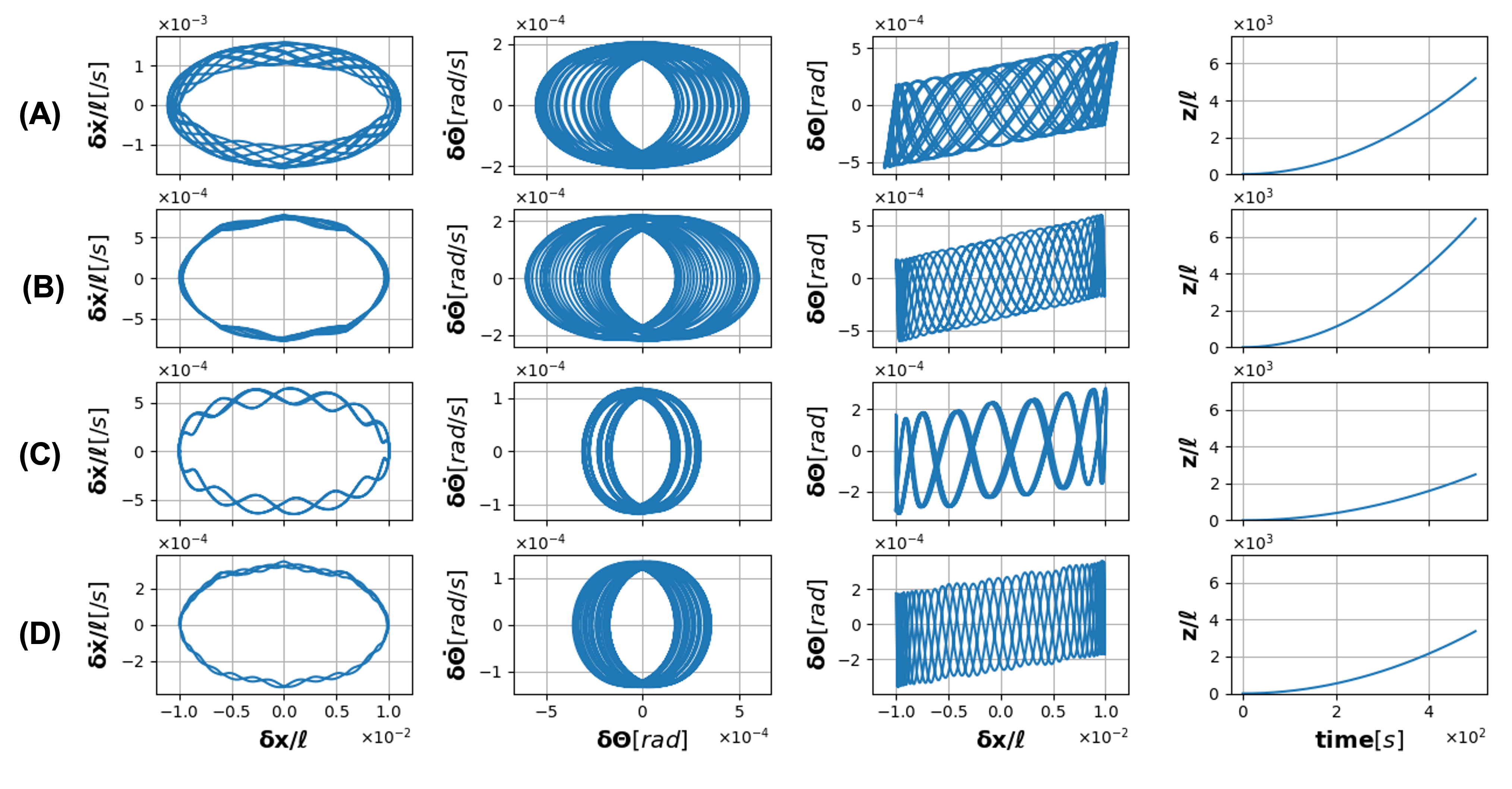}
\caption[]{Runge-Kutta solution of equation of motion for the two sail designs for both (beam-width, forward thrust):(A)($w_0=\ell/2, F_z/F_0=1.20$) (B)($w_0=\ell/2, F_z/F_0=1.70$) (C)($w_0=\ell, F_z/F_0=0.60$) (A)($w_0=\ell, F_z/F_0=0.45$)}
\label{fig:RKphasemap}
\end{figure}

The total force and torque on the bi-grating, however, is the result of the individual response of each panel with respect to the beam center. Thus, the force and torque are the functions of both angle and displacement from the equilibrium point. Shown in Fig.\ref{fig:Kmatrix} is the force and torque on the bi-grating for both the structures assuming an expanded filling ($w_0=\ell$) and non-expanded underfilling ($w_0=0.5\ell$) beam. It is evident from the figure the restoring force and torque are linear for small perturbations from the equilibrium point and the slope of these curves correspond to Jacobians as described in Eq.\ref{eq:linearEqnjacobian} for Linear stability analysis purposes.

Linear stability analysis however provides only a partial picture of stability i.e., the stability of the system for very small perturbations near the equilibrium point. And hence the equations of motion must be solved using numerical methods like Runge-Kutta to fully gain an understanding of system stability. For an initial perturbation of $(\delta x/\ell = 0.01, \delta \theta = 0.01 \mathrm{rad}, \delta \Dot{x}/\ell = 0, \delta \Dot{\theta} = 0)$, the equations of motion are solved for the two geometries and two different beam width $w_0=0.5\ell$ and $w_0=\ell$. The results are shown in Fig.\ref{fig:RKphasemap}. The phase maps for all the cases are closed, suggesting a stable system. Please note, these simulations were performed for $P_0=10$kW and for a gigawatt class laser as proposed in Breakthrough Starshot system will remain closed and bounded albeit the frequency of oscillation will be scaled accordingly. 

\begin{figure} 
\centering\includegraphics[scale=0.6]{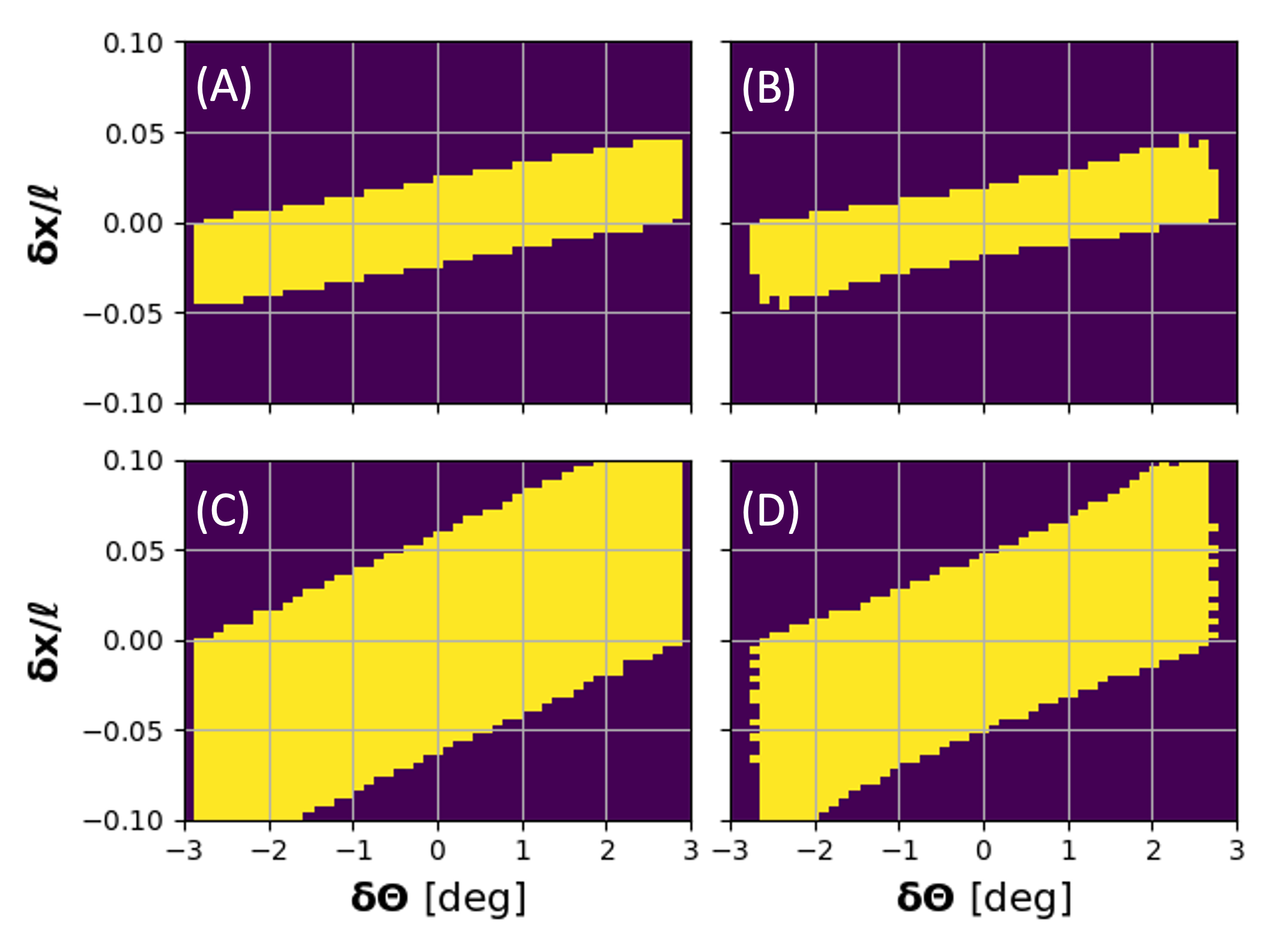}
\caption[]{Light Sail with stable trajectory for initital perturbation and rotation assuming (beam width, forward thrust): (A)($w_0=\ell/2, F_z/F_0=1.20$) (B)($w_0=\ell/2, F_z/F_0=1.70, 499/2601$) (C)($w_0=\ell, F_z/F_0=0.60$) (D)($w_0=\ell, F_z/F_0=0.45$)}
\label{fig:RKheatmap}
\end{figure}

The above simulation is repeated for a range of initial conditions to test the limits of stability and is shown in Fig.\ref{fig:RKheatmap}. It is evident from Fig.\ref{fig:RKheatmap}(A) and (B), an increase in forward thrust from 120\% to 170\% leads to 3.4\% shrinkage in stable basin for an underfilling beam-width. Similar behavior may be observed for an expanding beam Fig.\ref{fig:RKheatmap}(C) and (D) i.e., high forward thrust implies sacrifice on permissible stable conditions. However, the expanded beam imparts lower forward thrust than the non-expanded underfilling beam.



\section{Conclusion}
In conclusion, we designed, optimized, and cross-validated a rigid bi-grating light sail for stability and thrust. We proposed two different geometries with two different forward thrust forces while quantifying the bounds of stable initial conditions. The sail is stable for both underfilling and expanded filling beams. The design thus is an excellent choice for applications such as levitated optomechanics of millimeter/centimeter-scale objects or laser propelled meter scaled light sail for space missions limited to the solar systems. As a final remark, significant practical challenges exist in the realization of these sails. For example,  the design is very sensitive to wavelength and becomes unstable for Doppler-shifted wavelength in the case of relativistic sails of Breakthrough Starshot. If a tunable phased-array laser becomes a reality, the proposed design becomes an ideal choice. Moreover, the non-rigidity of the light sail must be considered and modeled using Lagrangian mechanics \cite{MODI1974}. Similarly, the local deformation of the all-dielectric thin sail may be optimized and modeled using the principles of conformal metasurfaces \cite{Wu2020}.


\begin{backmatter}

\bmsection{Funding}NASA Headquarters (80NSSC19K0975); Breakthrough Prize Foundation (7dBPF Starshot LLC).


\bmsection{Disclosures} The authors declare no conflicts of interest.

\bmsection{Data availability} Data underlying the results presented in this paper are not publicly available at this time but may be obtained from the authors upon reasonable request.

\end{backmatter}

\bibliography{Self-Stabilize}
\end{document}